\documentclass{ws-ijmpe}

\begin{document}

\markboth{Hasnaoui, Chomaz, Gulminelli}{Fermionic Molecular Dynamics for nuclear dynamics and thermodynamics}

%%%%%%%%%%%%%%%%%%%%% Publisher's Area please ignore %%%%%%%%%%%%%%%
\catchline{}{}{}{}{}
%%%%%%%%%%%%%%%%%%%%%%%%%%%%%%%%%%%%%%%%%%%%%%%%%%%%%%%%%%%%%%%%%%%%

\title{Fermionic Molecular Dynamics\\ for nuclear dynamics and thermodynamics
}

\author{\footnotesize K.H.O. Hasnaoui \& Ph Chomaz\footnote{Present address: IRFU/CEA Saclay}}

\address{Grand Acc\'el\'erateur National d'Ions Lourds (DSM-CEA/IN2P3-CNRS),\\ B.P.5027, F-14076 Caen cédex 5, France}

\author{F. Gulminelli}

\address{Laboratoire de Physique Corpusculaire de Caen (IN2P3-CNRS/Ensicaen et Universit\'e),\\ F-14050 Caen cédex, France}

\maketitle

\begin{history}
\received{(received date)}
\revised{(revised date)}
%\accepted{(Day Month Year)}
%\comby{(xxxxxxxxxx)}
\end{history}

\begin{abstract}
A new Fermionic Molecular Dynamics (FMD) model based on a Skyrme functional is proposed in this paper. After introducing the basic formalism,
 some first applications to nuclear structure and nuclear thermodynamics are presented.
\end{abstract}

\section{Introduction : the Fermionic Molecular Dynamics model}
The main goal of the Fermionic Molecular Dynamics model proposed by S. Dro\.zd\.z \textit{et al.}\cite{1} and B. Caurier \textit{et al.}\cite{2}, and improved thereafter by H. Feldmeier\cite{3}, is to modelise the dynamics of a fermions system where antisymetrisation is exactly taken into account. The FMD model consists of a single Slater determinant parametrised by a set of dynamical variational parameters $Q(t)=\left[q_{\mu}(t)| \mu =1,2,\ldots\right]$ :
\begin{eqnarray}
\vert Q(t) \rangle= \frac{\hat{\cal{A}}}{A!}\prod_{k}^{A}\vert q_{k}(t)\rangle 
\end{eqnarray}
where the single particle states are chosen to be gaussians, in order to localise the particles in phase space and be able to deal with particle density fluctuations and clustering:
\begin{eqnarray}
\left\langle \vec{r} \vert q_{k}(t) \right\rangle=\exp\left(-\frac{\left(\vec{r}-\vec{b}_{k}(t)\right)^{2}}{2a_{k}(t)}\right)\vert \chi_{k}(t), \phi_{k}(t) \rangle \vert m_{t}(k) \rangle
\end{eqnarray}
The real parameters $\chi_{k}$ and $\phi_{k}$ are the dynamical phases for the evolution of the spin degree of freedom, $\vert m(k) \rangle$ is the isospin degree of freedom, and the complex parameters $a_{k}$ and $\vec{b}_{k}$ are linked to the classical coordinates of nucleons. Indeed the expectation value of the position and the momentum for a single particle state reads :
\begin{eqnarray}
<\hat{\vec{r}}>&=&\vec{r}_{k}(t)=\text{Re}\left(\vec{b}_{k}(t)\right)+\frac{\text{Im}\left(a_{k}(t)\right)}{\text{Re}\left(a_{k}(t)\right)}\text{Im}\left(\vec{b}_{k}(t)\right)\\ <\hat{\vec{p}}>&=&\vec{p}_{k}(t)=\frac{\text{Im}\left(\vec{b}_{k}(t)\right)}{\text{Re}\left(a_{k}(t)\right)} 
\end{eqnarray}
The dynamical evolution of each variational parameters is obtained imposing a minimisation of the action :
\begin{eqnarray}
\delta\int_{t_1}^{t_2}dt\langle Q(t) \vert i\hbar\frac{d}{dt}-\hat{H}\vert Q(t)\rangle=0 
\end{eqnarray}
where $\hat{H}$ is the Hamiltonian of the system. This equation gives the FMD equation of motion for each variational parameter :
\begin{eqnarray}
\dot{q}_{\mu}=-\sum_{\nu}\mathcal{A}_{\mu\nu}^{-1}\frac{\partial \mathcal{H}}{\partial q_{\nu}} 
\end{eqnarray}
where $\mathcal{H}=\langle Q(t) \vert\hat{H}\vert Q(t)\rangle$.
An important feature of the FMD model is that the set of dynamical variational parameters includes the widths of the single particle states. 
This means that the quantum-mechanical spreading of a free wave packet can be reproduced by the model.
Another important point is that the single Slater approximation implies that the expectation value of the energy is given by 
a functional of the one body-density only, namely the Hartree-Fock energy $\langle Q(t) \vert\hat{H}\vert Q(t)\rangle=E_{HF}[f(\hat{\rho})]$. 
The Hilbert space is restricted respect to TDHF because of the gaussian wave packet constraint, meaning that FMD can be viewed as an approximation of TDHF. On the other side, it is interesting to remark that the FMD model gives the exact Hamilton equations for classical particles, meaning that all the correlations and fluctuations are taken into account at the classical level. 
Since the expectation value of the energy for the FMD model can be written as a density functional, the lastest Skyrme functionals as the SLy parametrisations\cite{4} can be used, with appropriate restoration of Galilean invariance\cite{5}.

\section{FMD for nuclear structure}
Our FMD model based on a Skyrme functional can be tested comparing observables with HF calculations and experimental data. 
In the following we show calculations of ground state properties and Giant Monopole Resonance (GMR) frequencies.
Concerning ground states, the FMD variational principle coincides with the minimization of the energy $\frac{\partial \mathcal{H}}{\partial q_{\nu}}=0$.
Then FMD becomes a static model where $\dot{q}_{\mu}=0$. To obtain the ground state properites, we have used an iterative algorithm based on a gradient method of first order. Table 1 represents the energies of the ground states for some light even-even nuclei. We can see that the results given by the FMD model with the SLy4 parametrisation are comparable to full HF calculations, and that both are close to the experimental data.
\begin{table}[!ht]
\begin{center}
 \begin{tabular}{|l|c|c|c|}
\hline $E_{GS}$ (MeV) & FMD & HF\cite{6} & Exp\cite{7} \\
\hline $^{4}$He & -26.156 & -26.7 & -28.296 \\
\hline $^{6}$Li & -30.225 & -32.478 & -31.99 \\
\hline $^{8}$Be & -44.996 & -45.439  & -56.5 \\
\hline $^{12}$C & -77.4 & -90.578  & -92.163 \\
\hline $^{16}$O & -127.9 & -128.485  & -127.62 \\
\hline $^{40}$Ca & -344.98 & -344.22  & -342.056 \\
\hline
\end{tabular}
\caption{FMD ground state energy with the SLy4 interaction for some even-even nuclei in comparison with results from the Hartree-Fock model, and  experimental data.}
\end{center}
\end{table}
A similar quality of results is obtained for mean square radii.
A discrepancy is however observed for $^{12}$C where FMD underestimates the binding energy. This can be tentatively understood observing that the FMD result has a pronounced $\alpha$ structure; similar states is experimentally observed as excited states.
Another interesting observable is the GMR frequency, which is a stringent test of the compressibility properties of nuclear matter. To excite a Giant Resonances, at a given time an exciting operator is applied to the system according to :
\begin{equation}
\hat{H}=\hat{H}_{\text{Nuclei}}+\lambda\hat{O}\delta(t)
\end{equation}
where $\hat{O}=\hat{\vec{r}}^{2}$ for the GMR, which characterizes the isotropic compression of the system at the initial time. In calculating the GMR for $^{12}$C with the SLy5 parametrisation, we found that the frequency of the monopole is given by $\omega_{FMD}=24.017$MeV, which is comparable to the experimental value $\omega_{Exp}=21.9\pm 0.3$MeV extracted by D.H. Youngblood {\it et al.}\cite{8}. Calculations for heavier systems and different multipolarities are in progress.

\newpage

\section{FMD for thermodynamics}
We have seen that the FMD model gives a good description of light systems at T=0MeV, ie good ground state properties and compressibility.
 We will propose now a method to extract the thermal properties at finite temperature.

The extraction of the phase diagram at finite temperature for a finite system needs 
the calculation of thermal averages of observable $<\mathcal{A}>_{\circ}$, as well as of the temperature $T$. If the system is ergodic, the thermal average is given by the time average $<\mathcal{A}>_{\circ}=\lim_{t\rightarrow\infty}\frac{1}{t}\int_{0}^{t}\mathcal{A}(t')dt'$. To know the temperature of the system, this latter has to be coupled to a thermometer, if we want the microcanonical temperature, or to a thermal bath if we want the canonical temperature. In both cases the coupling must be such that the total wave function is a tensor product between the nuclear system and the thermometer $|\Psi_{\text{Total}}\rangle=|\Psi_{\text{Nuclear}}\rangle\otimes|\Psi_{\text{Th}}\rangle $, where the total Hamiltonian is given by :
\begin{eqnarray}
 \hat{H}_{\text{Total}}=\hat{H}_{\text{Nuclear}}+\hat{H}_{\text{Th}}+\hat{H}_{\text{I}}
\end{eqnarray}
Here, $\hat{H}_{\text{I}}$ is the weak coupling between the two systems where the expectation values should  respect the conditions $<\mathcal{H}_{\text{Nuclear}}>\gg <\mathcal{H}_{\text{I}}>$ and $<\mathcal{H}_{\text{Th}}>\gg <\mathcal{H}_{I}>$. If the statistical ensemble is the microcanonical ensemble (thermometer case), we should always respect the condition $<\mathcal{H}_{\text{Nuclear}}>\gg <\mathcal{H}_{\text{Th}}>$, while for the canonical ensemble (coupling to a thermal bath) we should have $<\mathcal{H}_{\text{Th}}>\gg <\mathcal{H}_{\text{Nuclear}}>$. If the thermal eresponse of the thermometer (or the thermal bath) $<\mathcal{H}_{\text{Th}}>_{\circ}=f(T)$ is known analytically, it can be used to extract the common temperature of the two systems at equilibrium from the calculated time average $<\mathcal{H}_{\text{Th}}>_{\circ}$.

\begin{figure}[!ht]
\begin{center}
\mbox{
\epsfxsize=8cm
\epsfbox{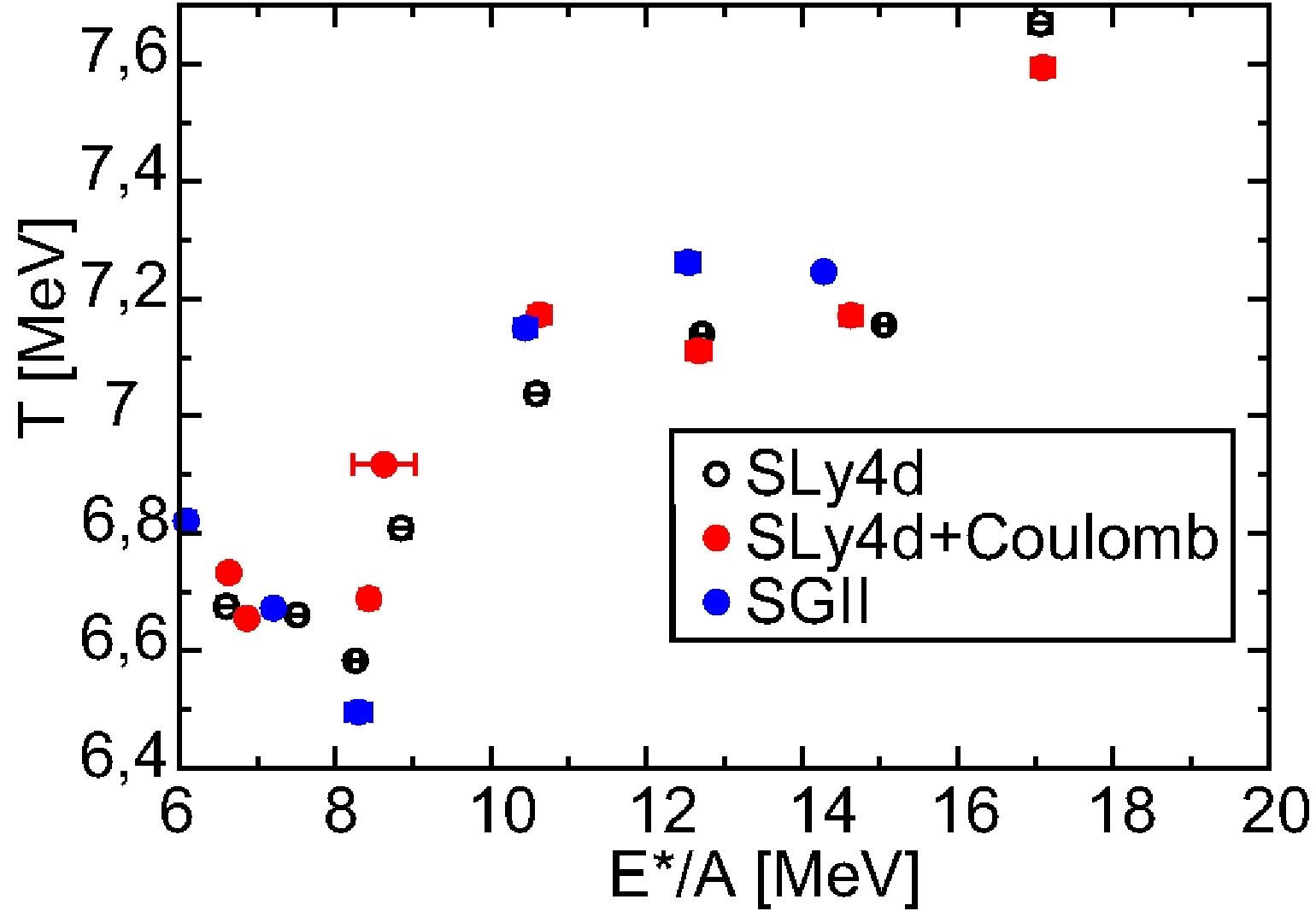}
}
\end{center}
\caption{Caloric curves for $^{4}$He with $W_{4}=0.01\text{MeV}.\text{fm}^{-4}$. The black, red, and blue dots represent respectively the calculations with the SLy4d interaction, SLy4d + Coulomb interaction, and the SGII interaction.}
\label{FIG:calorique}
\end{figure}

In practical calculations, the nucleus is trapped in a potential, and the total Hamiltonian for the nuclear system can be written as :
\begin{eqnarray}
\hat{H}_{\text{Nuclear}}=\hat{H}_{\text{Kinetic}}+\hat{H}_{\text{Interaction}}+\hat{H}_{\text{Trap}}
\end{eqnarray}
For the trap we have chosen a $\vec{r}^{4}$ potential such as $\hat{H}_{\text{Trap}}=W_{4}\sum_{i=1}^{A}\hat{\vec{r}}_{i}^{4}$
where $W_{4}$ is the trap constant. The thermometer has been taken as an harmonic oscillator where $\hat{H}_{\text{Th}}=\frac{\hat{\vec{p}}_{\text{th}}^{2}}{2m_{\text{th}}}+\frac{1}{2}m_{th}\omega_{\text{th}}\hat{\vec{r}}_{\text{th}}^{2}$,  
with thermal response $T=\hbar\omega/\ln\left(\frac{<\mathcal{H}_{\text{Th}}>_{\circ}+\frac{3}{2}\hbar\omega}{<\mathcal{H}_{\text{Th}}>_{\circ}-\frac{3}{2}\hbar\omega}\right)$. The coupling between the two system has been chosen as a contact interaction $\hat{H}_{\text{I}}=\lambda\sum_{i=1}^{A}\delta\left(\hat{\vec{r}}_{\text{th}}-\hat{\vec{r}}_{i}\right)$. Even if the parameters $W_{4}$,$\lambda$ and $\hbar\omega$ are chosen to optimise the thermalisation properties, the equilibration time is very long, of the order of 50000 fm/c. For this reason we have made calculations only for a very small system, namely $^{4}$He.
Applying the method explained above at different total energies, we have extracted the temperature as a function of the excitation energy per nucleon, the so-called caloric curve. Figures \ref{FIG:calorique} represent three calculations for a $^{4}$He nucleus using an hybrid ensemble intermediate between the microcanonical and the canonical ensemble, with $<\mathcal{H}_{\text{Th}}>\approx <\mathcal{H}_{\text{Nuclear}}>$. The caloric curves obtained with the SLy4d and the SGII parametrisation, with and without Coulomb interaction are presented. It is important to remark that the excitation energy axis contains the constraining potential and as such it is not a direct measure of the deposited energy.  
At low excitation energies the trend is not clear, and doubts may be shed on the ergodicity of the dynamics. At higher energy the curves show the expected trend for a transition from an excited fermion system to a system of non-interacting classical particles. This can be further confirmed
looking at the energy distribution, which is bimodal in the transition region.
Comparing the three caloric curves, we can see that that the choice of the nuclear interaction has small effects, and the same is true for the 
influence of Coulomb. Calculations with heavier systems are in progress.
\section*{Acknowledgements}

This work is supported by the NExEN ANR project.

\end{document}